\documentclass[11pt,a4,twoside,longbibliography]{article}
\usepackage[dvips,a4paper,left=2.0cm,right=2.0cm,top=2.0cm,bottom=2.0cm,headsep=1em]{geometry}
\usepackage{fancyhdr}
\usepackage{titlesec}
\usepackage[latin1]{inputenc}
\usepackage{amsmath,amsfonts,amssymb,array,amsthm,mathrsfs}
\usepackage{rotating}
\usepackage[font={small}]{caption}
\usepackage{lmodern,slantsc}
\usepackage{multirow}
\usepackage{chngcntr}
\usepackage{placeins}
\usepackage{caption}
\usepackage{enumitem}
\usepackage{verbatim}
\usepackage{cite}
\usepackage[sort&compress,numbers]{natbib}
\def\bibsep{6pt}
\def\bibfont{\small}
\usepackage[breaklinks, bookmarks=true,
colorlinks=true, linkcolor=blue, citecolor=blue, urlcolor=blue, filecolor=blue,
pdfborder={0 0 0}, pdfpagelabels]{hyperref}
\let\oldabstract\abstract
\let\oldendabstract\endabstract
\makeatletter
\renewenvironment{abstract}
{%
{\list{}{\addtolength{\leftmargin}{-0.2em} 
\listparindent 1.5em%
\itemindent\listparindent%
\rightmargin\leftmargin%
\parsep\z@\@plus\p@}%
\item\relax}%
{\endlist}%
\oldabstract}
{\oldendabstract}
\makeatother

\def\v0{\boldsymbol{0}}

\def\L{\langle}
\def\R{\rangle}
\newlength{\FigureHeight}
\newlength{\FigureHeightHalf}

\numberwithin{equation}{section}
\titleformat{\section}
{\large\bfseries}{\thetitle.}{0.5em}{}
\titleformat{\subsection}
{\normalfont\bfseries}{\thetitle.}{0.5em}{}
\titleformat{\subsubsection}
{\normalfont\itshape}{\normalfont\thetitle.}{0.5em}{}
\begin{document}

\title{An example from turbulence how not to use\\ the invariant function method of Lie-group symmetries}

\author{Michael Frewer$\,^1$\thanks{Email address for correspondence:
frewer.science@gmail.com}$\:\,$ \& George Khujadze$\,^2$ \\[1.0em]
\small $^1$ Heidelberg, Germany\\
\small $^2$ Chair of Fluid Mechanics, Universit\"at Siegen, 57068
Siegen, Germany}
\date{{\small\today}}
\clearpage \maketitle \thispagestyle{empty}

\begin{abstract}

\noindent
The recent Reply by Oberlack {\it et al.} [\href{https://doi.org/10.1103/PhysRevLett.130.069403}{Phys.~Rev.~Lett.~{\bf 130},~069403~(2023)}]\vphantom{\cite{Oberlack23}}$\,$\footnote{If interested in our response to the journal to a reply prior to their published one, see the \href{https://doi.org/10.5281/zenodo.7626464}{PRL-Rebuttal}.} fails to rebut the\linebreak[4] critique \cite{Frewer23,Brethouwer23} that a mathematical solution method has been misapplied in their original work \cite{Oberlack22}.\linebreak[4] On a point-by-point basis we prove that all arguments put forward in that Reply are incorrect.\linebreak[4] Therefore, the fact that the invariant solution method of Lie-group symmetries should not be used for unclosed systems in the same way as for closed systems still holds true. Ignoring this fact only leads to\linebreak[4] wrong conclusions. Claims such as having derived solutions of the statistical Navier-Stokes equations from first principles, or having found a measure for the intermittent behaviour of turbulence or particularly the true scaling in wall-bounded turbulent shear flows, are incorrect.

\vspace{0.5em}\noindent{\footnotesize{\bf Keywords:} {\it $\!$Statistical Physics, Turbulence, Shear Flow, Symmetry Analysis, Lie-Groups, Closure Problem}}
\end{abstract}

\section{On the DNS uncertainty of the fluctuation moments\label{Sec1}}

The results shown in Fig.$\,$4 in \cite{Oberlack23} are based on false assumptions. It is not true that the uncertainties of $R_3$ are two orders of magnitude above the values of $R_3$, as shown in Fig.$\,$4,
implying {\it ``that the uncertainties are so large that a valid statement about $R_3$ is meaningless"}.

The key mistake in \cite{Oberlack23} is the assumption that the uncertainty of the fluctuation moment $R_3$ can be determined deterministically from the given uncertainties of the full-field moments $H_i$,
in such a way\linebreak[4] that it is further assumed that these uncertainties are all independent of each other. But in turbulence, as is well known, such assumptions are not valid. They unaviodably lead to a wrong overestimation for the uncertainty of the fluctuation moments.

To yield a correct estimation, a stochastic method has to be used for the following simple reasons:\linebreak[4]
First, the full-field moments $H_i$ are stochastic quantities constructed or measured at each point in~space by sampling over time, making it possible that unfavorable uncertainties can cancel over time, where thus extreme unfavorable uncertainties can turn into statistical outliers. Second, since the $H_i$ are stochastic quantities, also their uncertainties~$\delta H_i$ are stochastic and therefore never
certain. And third, and most importantly, the construction or measurement of $H_1$, $H_2$, $H_3$, etc., are not independent of each other, but depend all on a single field, the full instantaneous streamwise
velocity $U:=U_1$, in the following~way:\footnote{The brackets $\L\cdot\R$ in \eqref{230119:1234} denote some averaging procedure.}
\begin{equation}
H_1=\big\L U\big\R,\qquad H_2=\big\L U^2\big\R,\qquad H_3=\big\L U^3\big\R,\quad \cdots
\label{230119:1234}
\end{equation}
that is, any uncertainty in the construction of $H_1$ will influence the uncertainty when constructing $H_2$
and so on. Respecting this fact, namely that the uncertainties of the full-field moments $H_i$ at each point are finely tuned coupled quantities constructed over a long sampling interval from a single field~$U$,\linebreak[4] will lead to verifiably meaningful results for the fluctuation moments when constructing them from the full-field moments, and not to a result that dramatically and falsely overestimates their uncertainty, as shown in Fig.$\,$4 in \cite{Oberlack23} for the third fluctuation moment $R_3$.

$\!$Instead of having the wrong relative uncertainty of more than 3650\%, as shown in Fig.$\,$4, for example\linebreak[4] at point~$x_2^+\approx 1000$, a realistic estimation based on a stochastic model (\cite{Frewer22}, Sec.$\,$C) shows that with a confidence level of 95\% the relative uncertainty of $R_3$ at that point will not grow beyond 10\% and with a confidence of even 99\% that it will not grow beyond 12\%.

This leads to the conclusion that in a DNS, if sampled over a sufficiently long time, the chances are almost certain that the sampled values $H_i$ have favorable uncertainties so that the fluctuation moment~$R_3$ can be determined with confidence, either indirectly via $R_3=H_3-3H_1H_2+2H_1^3$, or directly via $R_3=\L(U-H_1)^3\R$,
where both evaluations yield the very same result (\cite{Frewer22}, Sec.$\,$C).

\subsection{The consequence if the uncertainty of \texorpdfstring{$\boldsymbol{R_3}$}{R3} proposed by Oberlack \emph{et al.} were true}

Let's imagine for a moment that the uncertainty of the fluctuation correlation $R_3$ is as large as shown in Fig.$\,$4 in \cite{Oberlack23}. Then it would turn turbulence research upside down.

For instance, all studies that determine the degree of intermittency and anomalous scaling in shear flows with a mean velocity~$\bar{U}$, as in channel flow, would all turn invalid by this result shown in Fig.$\,$4.
For example, the study \cite{Toschi99}, where the authors then have to be told that their research in simulating the fluctuation correlations up to 7th order in channel flow is all wrong due to the `new result' shown by Oberlack {\it et al.} in Fig.$\,$4. To note here is that what the authors in \cite{Toschi99} are calling structure functions in their Eq.$\,$(1),
\begin{equation}
S_p(r^+,y^+)=\big\L\big\vert v_x(x^++r^+,y^+,z^+)-v_x(x^+,y^+,z^+)\vert^p\big\R,
\label{221203:2133}
\end{equation}
where $v_x$ represents the full instantaneous streamwise velocity, are indeed fluctuation correlations up to order~$p$, since the mean field in statistically stationary channel flow,
\begin{equation}
\bar{v}_x:=\big\L v_x(x^+,y^+,z^+)\big\R,
\end{equation}
is only a function of the wall-normal coordinate $y$, and therefore, when Reynolds decomposing the full-field $v_x$ into its mean and fluctuating part
\begin{equation}
v_x(x^+,y^+,z^+)=\bar{v}_x(y^+)+v^\prime_x(x^+,y^+,z^+),
\end{equation}
the mean field simply cancels out when taking the difference in \eqref{221203:2133}, thus leaving a pure fluctuation moment of $p$-th order
\begin{equation}
S_p(r^+,y^+)=\big\L\big\vert v^\prime_x(x^++r^+,y^+,z^+)-v^\prime_x(x^+,y^+,z^+)\vert^p\big\R.
\label{221203:2140}
\end{equation}

\vspace{0.25em}\noindent
Now, as \cite{Toschi99} results in the finding {\it ``that near the center of the channel the values $\zeta_p/\zeta_3$ [exponents~of~$S_p$]\linebreak[4] up to $p=7$ are consistent with the assumption of homogeneous and isotropic turbulence"} [p.$\,$1], it gives confidence enough that meaningful results for the fluctuation correlations up to 7th order were obtained, thus again clearly invalidating the arguments of Oberlack \emph{et al.}

Also, all research where higher-order fluctuation moments get determined by DNS or experiment, either to compare the data to turbulence models or to assess the validity of scaling laws, as in
\cite{Marusic13,Wilczek14,Smits15,Jeyapaul15,Brethouwer18}, for example, would all have to be cancelled if the new result of Oberlack {\it et al.} were true, because with the uncertainty of the fluctuation moments being so absurdly large, no results that involve higher-order fluctuation moments can be trusted anymore.

In particular, by claiming {\it ``that the uncertainties are so large that a valid statement about $R_3$ is meaningless"} and that {\it ``comparing DNS data with $R$-moments for $n>2$ is meaningless"}
\cite[p.$\,$2]{Oberlack23}, which, if true, would invalidate a great number of papers. But, of course, this is not the case, because the arguments of Oberlack {\it et al.} are simply based on wrong reasoning.

No research paper that carefully determines the higher-order fluctuation moments $R_{n}$ from DNS or experiment, either directly, via evaluating $\L (U-\bar{U})^n\R$, or indirectly, via expanding this expression in terms of the full-field moments $H_n=\L U^n\R$, is in `danger' by Oberlack's `new' result.

\newgeometry{left=2.0cm,right=2.0cm,top=2.0cm,bottom=1.5cm,headsep=1em,footskip=0em}

\section{On the indicator functions to test scaling}

The comparison shown in Fig.$\,$1 in \cite{Oberlack23} is invalid and highly misleading. The indicator function used in \cite{Oberlack23} to generate Fig.$\,$1, namely Eq.$\,$(18) in \cite{Oberlack22},
\vspace{-0.5em}
\begin{equation}
\Gamma_n=\frac{y^+}{\overline{U^n_1}^{\,+}+B_n}\frac{d\overline{U^n_1}^{\,+}}{dy^+},
\label{220121:1122}
\end{equation}
is not part of the class of indicator functions standardly used to test data on power- or log-law scaling in turbulence, which, respectively, are
\vspace{-0.75em}
\begin{equation}
\Gamma=\frac{y^+}{\mathcal{F}}\frac{d\mathcal{F}}{dy^+},\qquad \Xi=y^+\frac{d\mathcal{F}}{dy^+},
\label{220121:1123}
\end{equation}
where $\mathcal{F}$ is some statistical correlation function that can be built from the data alone, i.e., a function\linebreak[4] that should not involve any modelling parameters. Obviously, the power-law indicator
function $\Gamma_n\!$~\eqref{220121:1122}\linebreak[4] is not of this type \eqref{220121:1123}, simply because the defining correlation functions $\mathcal{F}_n$ associated to $\Gamma_n$,
\begin{equation}
\Gamma_n\,=\,\frac{y^+}{\overline{U^n_1}^{\,+}+B_n}\frac{d\overline{U^n_1}^{\,+}}{dy^+}\;\equiv\;\frac{y^+}{\mathcal{F}_n}\frac{d\mathcal{F}_n}{dy^+},\quad\;\text{where}\quad\mathcal{F}_n=\overline{U^n_1}^{\,+}+B_n,
\end{equation}
are not functions of the data $\overline{U^n_1}^{\,+}$ alone, but depend on the modelling parameters $B_n$. In other words,\linebreak[4] the indicator function $\Gamma_n$~\eqref{220121:1122} is {\it not} an unbiased indicator function, as the standard ones \eqref{220121:1123} are.\linebreak[4]
By adding the modelling parameters $B_n$ to the data $\overline{U^n_1}^{\,+}\!$, it modifies the data in bias towards the modelling function used, which itself is \cite{Oberlack23,Oberlack22}
\begin{equation}
\overline{U^n_1}^{\,+}=C_n(y^+)^{\omega(n-1)}-B_n,
\label{230121:1318}
\end{equation}
making use of exactly these parameters $B_n$. Hence, the indicator function $\Gamma_n$~\eqref{220121:1122} is a biased indicator function that, by construction, automatically will show a good match to DNS data only for its own specifically designed modelling function \eqref{230121:1318}, while for all other modelling functions it will predictably show a poor match. Therefore, it is not surprising to see a mismatch in Fig.$\,$1 in \cite{Oberlack23}, simply because it is based on an unfair comparison and thus also pointless to show.

To make a fair comparison of the matching quality between the scaling law \eqref{230121:1318} and the $n$-th power of the mean velocity $\bar{U}_1$, one has two choices: Either a biased (model-dependent) indicator function for each functional class is used, which, of course, is not so indicative and conclusive since the prediction is biased towards the modelling function, or, better, a common unbiased (model-free) indicator function is used for both functional classes.

Such a fair comparison is shown in \cite[Sec.$\,$B.1]{Frewer22}, demonstrating that for both approaches the matching of the $n$-th power of the mean velocity to the DNS data is of the {\it same} quality as that of the scaling law \eqref{230121:1318}. This clearly invalidates Fig.$\,$1 in \cite{Oberlack23} and its conclusion that {\it ``the $n$-th power of the mean velocity in figure 1 apparently is a poor approximation"}, because just the opposite can be shown, if only fair comparisons are made.

\section{On the fitting result of \texorpdfstring{$\boldsymbol{R_2}$}{R2} in the inertial layer}

\vspace{-0.5em}
In Fig.$\,$3 in \cite{Oberlack23}, it is shown how well their scaling law for the second fluctuation moment $R_2=\overline{u^2_1}^{\,+}\!$ agrees with the DNS data. What is not discussed is how unstable this fit is. In particular, their scaling~law\linebreak[4]
\vspace{-0.75em}
\begin{equation}
\overline{u^2_1}^{\,+}=C_2(x_2^+)^\omega-B_2-\bar{U}_1^{+2},\;\;\text{with}\;\;\bar{U}^+_1=(1/\kappa)\ln(x_2^+)+B,
\label{230121:1803}
\end{equation}
is not robust to small changes in the fitted parameters, as it should be for a function that is fitted to data. A tiny change of only 0.1\% in one of the parameters $\omega$, $B_2$ or $C_2$, as shown in \cite[Sec.$\,$B.2]{Frewer22},\linebreak[4] already leads to large visible mismatch, e.g., in the region $10^3<x_2^+<3\cdot 10^3$ even up to more than 30\%.\linebreak[4]
The reason for this high instability of \eqref{230121:1803} lies in the unnatural large values of the shift and normalization parameters $B_2$ and $C_2$, which have to compensate for the unnatural term $\bar{U}_1^{+2}$
of a mean flow within a fluctuation moment.\footnote{Surely, for the mean flow $\bar{U}_1$
to appear in a modelled equation for a fluctuation moment is not unnatural, but for~it\linebreak[4] to explicitly appear in a solution to that equation is. Therefore, even from this perspective, the scaling result \eqref{230121:1803} turns out to be exceedingly unnatural, and thus, in itself, not conclusive to be a valid solution to a fluctuation~moment.} The values of $B_2$ and $C_2$ in the inertial range are of order 100 for a quantity $R_2$ \eqref{230121:1803} to match that varies only of order 1 in this range. Fig.$\,$3 in \cite{Oberlack23} therefore does not show a good fit of \eqref{230121:1803}. Instead, it shows a poor fit that only fits the data very unnaturally.

\restoregeometry

\section{On the trivial and universal scaling in channel center}

As analyzed and discussed in \cite[Sec.$\,$2]{Frewer22}, the scaling shown in Fig.$\,$2 in \cite{Oberlack23} is not a phenomenon of turbulence. It is a trivial and universal feature that can be obtained with any function $f$ that is non-zero at the origin (channel-center) and then plotted in deficit form in a log-log diagram along with its powers $f^n$. If the function $f$ starts off linearly at the origin, then the slope near the
origin is~1,\linebreak[4] if it starts off as a quadratic extremum, then the slope is 2, and so on, which can be readily seen when performing a Taylor expansion around the origin $x=0$ (channel center):
\begin{equation}
\left.
\begin{aligned}
\text{Slope-1 case}\!:\quad &
y=f^n-f_0^n,\quad f_0:=f(0)\neq0,\quad f_0^\prime:=f^\prime(0)\neq0, \quad n\geq 1\\[0.4em]
&\phantom{y}=(f_0+x\cdot f^\prime_0+\mathcal{O}(x^2))^n-f^n_0\\[0.4em]
&\phantom{y}=f^n_0+nf_0^{n-1}f_0^\prime\cdot x+\mathcal{O}(x^2)-f_0^n\\[0.4em]
&\phantom{y}=nf_0^{n-1}f_0^\prime\cdot x,\quad 0<x\ll 1 \\[0.4em]
&\hspace{-0.9cm}\Leftrightarrow\;\: Y=X+c_1,\quad\forall n\geq 1,\quad Y:=\log(y),\quad X:=\log(x),\quad c_1:=\log(nf_0^{n-1}f_0^\prime),
\\[1.0em]
\text{Slope-2 case}\!:\quad &
y=f^n-f_0^n,\quad f_0\neq0,\quad f_0^\prime=0,\quad f_0^{\prime\prime}\neq0,\quad n\geq 1\\[0.4em]
&\phantom{y}=(f_0+x^2\cdot {\textstyle\frac{1}{2}}f^{\prime\prime}_0+\mathcal{O}(x^3))^n-f^n_0\\[0.4em]
&\phantom{y}=f^n_0+{\textstyle\frac{n}{2}}f_0^{n-1}f_0^{\prime\prime}\cdot x^2+\mathcal{O}(x^3)-f_0^n\\[0.4em]
&\phantom{y}={\textstyle\frac{n}{2}}f_0^{n-1}f_0^{\prime\prime}\cdot x^2,\quad 0<x\ll 1 \\[0.4em]
&\hspace{-0.9cm}\Leftrightarrow\;\: Y=2X+c_2,\quad \forall n\geq1,\quad c_2:=\log({\textstyle\frac{n}{2}}f_0^{n-1}f_0^{\prime\prime}).
\end{aligned}
~~~\right\}
\label{230122:1611}
\end{equation}
The even correlation functions of $U_2$ and $U_3$, as shown in Fig.$\,$2 in \cite{Oberlack23}, are exactly of the above slope-2 type. The same is true for all moments of the full-field $U_1$ and all moments of the fluctuation field $u_1=U_1-\bar{U}_1$. Hence, it is not surprising that they all have the universal slope $2$ near channel center. And the fact that all slopes are so constantly long-tailed is only a feature of the log-log presentation used. The reason that for all these moments Oberlack {\it et al.} only measure a scaling of about 1.95, and not exactly 2, is that they simply take a too long fitting range away from channel center, which naturally leads to a deviation from pure quadratic scaling, since obviously higher-order Taylor terms become relevant, which will deviate from the straight line $Y\sim2X$ \eqref{230122:1611} in the log-log representation.

So, the fact that we get straight parallel lines with slope 2 near channel center for all the correlation functions mentioned above is nothing special and is trivial to predict from simple Taylor asymptotics.
No sophisticated invariant scaling theory is needed to predict this trivial and universal structure. However, if one still wishes to formulate this trivial scaling by an invariant scaling, then a consistent approach can be found in \cite[Sec.$\,$4.1]{Frewer22}.

As a side remark, interesting to note here is that Oberlack {\it et al.} only show the trivial channel-center scaling for diagonal moments, in \cite{Oberlack23} as well as in \cite{Oberlack22}. And it is clear why. Because, if they would show any off-diagonal moments, say $\overline{U^n_1U_2}$, their scaling theory will fail, although it should not, according to them, because their scaling theory does not restrict to any specific spatial direction or to any fixed length scale. Their generated scaling theory allows for any direction and length-scale in the~flow. In other words, their scaling theory always results in isotropic scaling, no matter how large the length scale is considered to be. But, of course, isotropic scaling on large length scales does not apply to flows such as channel flow, which are clearly anisotropic on large length scales.

In particular, the failure of their scaling theory to predict in channel center the scaling of the off-diagonal moments $\overline{U^n_1U_2}$ expresses itself in the fact that their scaling laws cannot go beyond parallel lines in the setting mentioned above. But, the off-diagonal moments $\overline{U^n_1U_2}$ are of a completely different structure. Due to the relation
$\overline{U^n_1U_2}=\overline{u^n_1u_2}+\sum_{k=1}^{n-1}\binom{n}{k}\bar{U}_1^k\overline{u_1^{n-k}u_2}$ they are zero at channel center (since~$\overline{u^n_1u_2}$ are all
odd functions), which means that a structure of parallel lines does not exist for this~case (analogous to what is shown in Fig.$\,$2 in \cite{Frewer22x}, namely a different slope for each $n$), and therefore cannot be predicted by their scaling theory, although it should, according to them, but in reality it does not.

\newgeometry{left=2.0cm,right=2.0cm,top=2.0cm,bottom=1.9cm,headsep=1em}

\section{Conclusion}

The reason why the study \cite{Oberlack22} is incorrect in all points it presents, and the subsequent Reply \cite{Oberlack23} incorrect in all points it discusses, is that a mathematical method is being misapplied here. A method developed for finding invariant solutions in {\it closed} systems is applied here one-to-one to a system that is unclosed. As shown in \cite{Frewer22}, such an approach leads to all kinds of mistakes and wrong conclusions if naively applied to the unclosed statistical equations of turbulence and no modelling is attempted.

For example, a key mistake in \cite{Oberlack22}, as well as in any other publication by Oberlack {\it et~al.} on this topic, is that from an infinite pool of possible invariant transformations always the same two invariant transformations are chosen, Eqs.$\,$(8)-(9) in \cite{Oberlack22}, that cannot be realized by the governing equations, neither by the Navier-Stokes nor by the Euler equations. The reason is that these two transformations violate the classical principle of cause and effect between the fluctuations and the mean fields~\cite{Frewer22,Frewer14.1,Frewer15,Frewer16.1,Frewer17,Frewer18.2,Sadeghi20},\linebreak[4] with the result that the induced invariant functions are not solutions of the governing equations.

Because, if the scaling laws of the full-field moments $H_i$ are solutions, as Oberlack {\it et~al.} in \cite{Oberlack22} claim to have, since these laws apparently solve the unclosed hierarchy of all moment equations, then they should also be solutions when mapped to the fluctuation moments $R_i$, which is done by a Reynolds decomposition, an analytical one-to-one map that does not change the solution property of a function. But this is not the
case here, as shown by us through the red lines in Figs.$\,$1(c-d) in \cite{Frewer23}. And this failure is structural and not due to a numerical stability issue, or any DNS uncertainty when transforming to the fluctuation moments, as proven in \cite[Sec.$\,$C]{Frewer22} and once again discussed herein in~Sec.$\,$\ref{Sec1}. This unambiguously proves that the proposed scaling laws in \cite{Oberlack23} are not solutions of the statistical Navier-Stokes equations, both in the inertial layer as well as in channel center.

As analytically and rigorously proven in \cite{Frewer14.1,Frewer15,Frewer17}, the inconsistency in the invariant functions in \cite{Oberlack22} already starts at moment order $n=2$, and then systematically infects all higher orders with increasing intensity. And exactly this is expressed as the matching failure shown in Figs.$\,$1(c-d) in~\cite{Frewer23}. While particularly in the log-layer for $n=2$ appropriate tools from statistical data analysis are still needed to detect this failure, the matching failure for the next higher order $n=3$ is already so pronounced in this layer that no further analysis is needed. As already said, the reason for this failure lies in the fact that the two arbitrarily chosen invariant transformations in \cite{Oberlack22}, Eqs.$\,$(8)-(9), are nonphysical in violating the classical principle of cause and effect and therefore cannot be realized by the governing equations being simulated. Discarding these invariances of \cite{Oberlack22} and replacing them by physical ones, improves the matching of the fluctuation moments to the data by several orders of magnitude, as shown e.g. in \cite[Sec.$\,$5]{Frewer14.1}, or \cite[Sec.$\,$4]{Frewer22}, which is a further
clear indication that these two\linebreak[4] statistical invariances in \cite{Oberlack22} are unphysical. To note is that these two unphysical invariances were also used to generate flawed and misleading scaling laws in other flow configurations, as channel flow with wall transpiration \cite[Sec.$\,$E]{Frewer22} or jet~flow~\cite{Sadeghi20}.

The failure of \cite{Oberlack23} explicitly shows what pitfalls lurk if the invariant function method of Lie-group symmetries is not applied correctly. The problem is the functionally unclosed nature of turbulence, which prevents the method from being applied as is standardly done to generate invariant solutions for functionally {\it closed} systems. Even if one formally considers all equations of the infinite and unclosed hierarchy of statistical equations and extracts from it an invariant transformation that does {\it not} violate the causality principle and other immutable constraints,\footnote{For instance, for the infinite and thus unclosed hierarchy of equations of the probability density function (PDF), there are five additional constraints that need to be respected by every invariant transformation of this system. These~are the so-called non-negativity, normalization, coincidence, separation and conditional constraints \cite{Frewer17,Frewer15}. In particular, the reviews \cite{Frewer21.2,Frewer21.3,Frewer21.6,Frewer21.4} explicitly show what happens if one of these constraints are violated.}
there still is no guarantee that the underlying governing (closed) equations will realize this invariant transformation,\footnote{Obviously, \hspace{-0.02mm}if \hspace{-0.02mm}an \hspace{-0.02mm}invariant \hspace{-0.02mm}transformation \hspace{-0.02mm}already \hspace{-0.02mm}violates \hspace{-0.02mm}from \hspace{-0.02mm}the \hspace{-0.02mm}outset \hspace{-0.02mm}one \hspace{-0.02mm}of \hspace{-0.02mm}the \hspace{-0.02mm}dynamical \hspace{-0.02mm}constraints \hspace{-0.02mm}as \hspace{-0.02mm}causality,\linebreak[4] as the two invariant transformations Eqs.$\,$(8)-(9) in \cite{Oberlack22} do, then, of course, it is 100\% certain that such an invariance will~not,~can not and never will be realized by the governing equations, neither globally nor locally.} despite leaving all equations of the infinite hierarchy invariant and reducing it to a $0=0$ identity through its associated invariant functions. The problem is that for unclosed systems the strong concept of symmetry turns into the weaker concept of equivalence, where solutions are not necessarily mapped to solutions anymore \cite{Frewer14.1,Frewer22}. This very basic but crucial fact is being ignored by M. Oberlack for more than two decades now, which since then has led to all kinds of misleading and flawed results.

\restoregeometry

\nocite{apsrev42Control}
\bibliographystyle{apsrev4-2}
\bibliography{References}

\end{document}